\documentclass[aps,prb,twocolumn,superscriptaddress]{revtex4}
\usepackage{ amssymb }
\usepackage{graphicx}
\usepackage{float}
\usepackage{subfigure}
\usepackage{booktabs}
\usepackage{array}
\usepackage{natbib}
\usepackage{collref}

\begin{document}


\title{Weak Topological Insulators in PbTe/SnTe Superlattices}




\author{Gang Yang}
\affiliation{Department of Physics, The Pennsylvania State University, State College, PA 16802, USA}
\affiliation{Department of Physics and State Key Laboratory of Low-Dimensional Quantum Physics, Tsinghua University, Beijing 100084, People's Republic of China}
\author{Junwei Liu}
\affiliation{Department of Physics and State Key Laboratory of Low-Dimensional Quantum Physics, Tsinghua University, Beijing 100084, People's Republic of China}
\author{Liang Fu}
\affiliation{Department of Physics, Massachusetts Institute of Technology, Cambridge, MA 02139, USA}
\author{Wenhui Duan}
\email{dwh@phys.tsinghua.edu.cn}
\affiliation{Department of Physics and State Key Laboratory of Low-Dimensional Quantum Physics, Tsinghua University, Beijing 100084, People's Republic of China}
\author{Chaoxing Liu}
\email{cxl56@psu.edu}
\affiliation{Department of Physics, The Pennsylvania State University, State College, PA 16802, USA}


\date{\today}

\begin{abstract}
It is desirable to realize topological phases in artificial structures by engineering electronic band structures. In this paper, we investigate (PbTe)$_m$(SnTe)$_{2n-m}$ superlattices along the [001] direction and find a robust weak topological insulator phase for a large variety of layer numbers $m$ and $2n-m$. We confirm this topologically non-trivial phase by calculating $Z_2$ topological invariants and topological surface states based on the first-principles calculations. We show that the folding of Brillouin zone due to the superlattice structure plays an essential role in inducing topologically non-trivial phases in this system. This mechanism can be generalized to other systems in which band inversion occurs at multiple momenta, and gives us a brand-new way to engineer topological materials in artificial structures.
\end{abstract}

\pacs{71.20.-b,73.20.-r,73.21.Cd,71.28.+d}

\maketitle

\section{Introduction}
The concept of time reversal (TR) invariant topological insulators (TIs) characterizes a new class of materials that are insulating in the interior of a sample, but whose surfaces contain robust conducting channels protected by TR symmetry.\cite{RevModPhys823045, RevModPhys831057,moore2010,qi2010phystoday} Since the metallic surface states of TIs can be described by Dirac fermions with exotic physical properties, researchers have been interested in pursuing different types of topological materials.\cite{RevModPhys823045, RevModPhys831057} TR invariant TIs can exist in both two dimensions and three dimensions. TIs in two dimensions, also known as ``quantum spin Hall insulators", were first predicted theoretically.\cite{PhysRevLett95226801,PhysRevLett96106802,Bernevig15122006} The quantum spin Hall effect was experimentally observed in HgTe/CdTe quantum wells (QWs)\cite{König02112007} and in InAs/GaSb type II QWs \cite{PhysRevLett107136603,PhysRevB81201301,2013arXiv13061925D}. In three dimensions, TR invariant TIs can be further classified into two categories, strong TIs and weak TIs.\cite{PhysRevLett98106803,PhysRevB76045302,PhysRevB75121306,PhysRevB79195322} Strong TIs have an odd number of Dirac cones at their surfaces and can be characterized by one $Z_2$ topological invariant, while the surface states of weak TIs contain an even number of Dirac cones and three additional $Z_2$ topological invariants are required.\cite{PhysRevB76045302} Strong TIs have been realized experimentally in various classes of materials, including Bi$_{1-x}$Sb$_{x}$, \cite{PhysRevB76045302,PhysRevB80085307,nature06843}  Bi$_{2}$Se$_{3}$ family,\cite{Nphys101038,Chen10072009,PhysRevLett103146401}  $\mathrm{TlBiTe_2}$ family,\cite{PhysRevLett105266401,PhysRevLett105136802}  $\mathrm{PbBi_{2}Se_{4}}$ family, \cite{2010arXiv10075111X,PhysRevB83041202} strained bulk HgTe,\cite{PhysRevLett98106803,PhysRevLett106126803,PhysRevLett107136803} {\it etc}. In contrast, only a few practical systems for weak TIs have been proposed theoretically \cite{PhysRevLett109116406,PhysRevB84075105,2013arXiv13078054T} and no experiments on weak TIs have been reported, up to our knowledge. Surface states of weak TIs are also expected to possess intriguing phenomena, such as one-dimensional helical modes along dislocation lines,\cite{ranyingnaturephys} the weak anti-localization effect,\cite{PhysRevLett108076804} the half quantum spin Hall effect,\cite{Liu2012906} {\it etc}.\cite{PhysRevB84035443,PhysRevB86245436,PhysRevB88045408,PhysRevB86045102,2012arXiv12126191F,PhysRevLett109246605}
Impurity scattering is reduced for surface states of TIs, so electric currents can flow with low dissipation. This leads to wide-ranging interests in device applications of TIs.

To search for new topological materials with robust physical properties, it is essential to engineer electronic band structures with the desired features. A useful and intuitive physical picture to understand TIs is the concept of ``band inversion''. Band inversion means that the band orderings of conduction and valence bands are changed at some high symmetry momenta in the Brillouin zone (BZ), so that band dispersion cannot be adiabatically connected to the atomic limit of the system under certain symmetries.\cite{PhysRevB79195321,PhysRevB76045302,Bernevig15122006} In other words, the band gap changes its sign from positive to negative when band inversion occurs (We usually define the sign of a normal band gap to be positive). Actually all the well-known topological materials have ``inverted'' band structures. A large variety of experimental methods can be applied to tune band gaps, such as controlling chemical compositions by doping,\cite{nature06843,nmat3449,PhysRevLett110206804} applying strain,\cite{PhysRevLett106126803} {\it etc}. In this paper, we propose that band inversion can be controlled by constructing superlattice structures, which has the obvious advantage of their controllability. We will consider the PbTe/SnTe superlattice, a typical semiconductor superlattice made of IV-VI group compounds, as an example due to its simplicity. We will show that weak TIs can be achieved for PbTe/SnTe superlattices grown along the [001] direction. Remarkably, the weak TI phase we found is \emph{not} equivalent to a stack of 2D quantum spin Hall layers in [001] direction. Instead, the nontrivial topology arises from the folding of BZ, which plays an essential role in inducing band inversion. This mechanism can be in principle generalized to search for other topological phases, including strong TIs and topological crystalline insulators, and paves the way to engineering topological phases in artificial structures.

This paper is organized in the following way. In section II, we briefly review the bulk properties of PbTe and SnTe, and introduce our calculation methods. In section III, we will show the evolution of band structure from PbTe to a $\mathrm{(PbTe)_{1}(SnTe)_{1}}$ superlattice step by step to illustrate how we obtain weak TIs. We calculate topological invariants, as well as surface states of the $\mathrm{(PbTe)_{1}(SnTe)_{1}}$ superlattice, to demonstrate the weak TI phase in this system. We will also discuss experimental realization and other possibilities to realize weak TI based on the same mechanism. The conclusion is drawn in section IV.


\section{Superlattice configuration and calculation methods}
We start with a review of material properties of bulk PbTe and SnTe. PbTe and SnTe have face-centered cubic NaCl-type of structures with the corresponding BZ shown in Fig. \ref{pic:fig1}. Both materials are narrow gap semiconductors with multiple applications in thermoelectricity, infrared diode and even superconductivity\cite{PhysRevB5513605,PhysRevB75195211,PhysRevB81245120}. The band gaps of these materials are located at the center of the hexagon on the BZ boundary, usually denoted as L points (See Fig. \ref{pic:fig1}(b)). There are four L points in totel, which are related to each other by  mirror symmetry. It was shown that the band gap of SnTe has the opposite sign of that of PbTe.\cite{ncomms101038} Consequently, the systems consisting of SnTe and PbTe may possess topologically non-trivial properties. For example, recent interest in these IV-VI semiconductors is stimulated by the prediction that SnTe represents a new type of topological phase dubbed ``topological crystalline insulators", which host gapless surface states protected by mirror symmetry. \cite{PhysRevLett106106802,ncomms101038,fang2012a,slager2012,jw} This finding was recently confirmed by the experimental observation of surface states in SnTe family of materials.\cite{nphys2442,xu2012a,nmat3449} With additional uniaxial strain along the [111] direction, both PbTe and SnTe are expected to be strong TIs.\cite{PhysRevB76045302,PhysRevLett105036404} A more recent theoretical work \cite{PhysRevB85205319} shows that, in a large thickness range, PbTe/SnTe superlattice along the [111] direction exhibit properties similar to a strong TI phase.  In this work, we will consider a superlattice with alternative stacks of SnTe and PbTe layers along the [001] direction, denoted as (PbTe)$_m$(SnTe)$_{2n-m}$, where $m$ and $2n-m$ represent the number of PbTe and SnTe layers, respectively. Fig. \ref{pic:fig1}(a) shows the (PbTe)$_1$(SnTe)$_1$ superlattice as an example.

The calculations are performed within the framework of density functional theory (DFT) calculations using the Perdew-Burke-Ernzerhof (PBE) generalized gradient approximation\cite{PhysRevLett773865} and the projector augmented wave (PAW) potential\cite{PhysRevB5017953}, as implemented in the Vienna \emph{ab initio} simulation package (VASP)\cite{PhysRevB5411169}. The spin-orbit coupling is included in all the calculations. The energy cutoff of the plane-wave basis is 340 eV. The 10$\times$10$\times$10 and 10$\times$10$\times$1 Monkhorst-Pack \emph{k} points are used for bulk and surface calculations separately. The lattice parameters are obtained by structural optimization.

Different topological phases can be determined by calculating $Z_2$ topological invariants.\cite{PhysRevB74195312,PhysRevB76045302} In three dimensions, there are four topological invariants, one strong topological index and three weak topological indices. With space inversion symmetry preserved, topological invariants can be evaluated easily using the parity of occupied states at eight time-reversal-invariant momenta (TRIM) $\Gamma_i$ ($i=1,\cdots,8$).\cite{PhysRevB76045302} The strong topological index $\nu_{0}$ is given by
\begin{equation}
\label{equ:strongTI}
(-1)^{\nu_{0}}=\prod_{i=1}^{8} \delta_{\Gamma_i},
\end{equation}
and the three weak topological indices $\nu_{k}$ are given by
\begin{equation}
\label{equ:weakTI}
(-1)^{\nu_{k}}=\prod _{n_{k}=1;n_{j\neq k}=0,1} \delta_{\Gamma_{i=(n_{1},n_{2},n_{3})}},
\end{equation}
where $\delta_{\Gamma_i}=\prod _{n \in occ} \xi _{2n}(\Gamma _{i})$ is the product of parity of all occupied states for one time-reversal copy at TRIM $\Gamma _{i}=\Gamma_{i=(n_{1},n_{2},n_{3})}=(n_{1}\mathrm{\vec{b}_{1}}+n_{2}\mathrm{\vec{b}_{2}}+n_{3}\mathrm{\vec{b}_{3}})/2 $ and $\mathrm{\vec{b}_{i}}$ are the reciprocal lattice vectors.\cite{PhysRevB76045302} A strong TI phase is determined by the index $\nu_0$ while a weak TI phase is characterized by a vector $\bar\nu=(\nu_1,\nu_2,\nu_3)$. Weak TIs can be understood as a stacking of 2D TI layers along the direction $\mathrm{\vec{G}_{\nu}=\nu_{1}\vec{b}_{1}+\nu_{2}\vec{b}_{2}+\nu_{3}\vec{b}_{3}}$.\cite{PhysRevB76045302}
On the surfaces with miller index $h \neq \bar{\nu}(mod2)$ in weak TIs, an even number of Dirac surface states can appear.\cite{PhysRevB76045302} As a check of our numerical methods, we calculate the parity of all occupied states at TRIM for the bulk SnTe and PbTe, as shown in Table \ref{table:parity}. As expected, both strong topological index and weak topological indices are trivial for SnTe and PbTe, which is consistent with the previous analysis.\cite{PhysRevB76045302} The underlying reason is that there are four L points with $\delta_{\Gamma_i}=+$ for SnTe and $\delta_{\Gamma_i}=-$ for PbTe ($\Gamma_i=\mathrm{L_{1,2,3,4}}$), as shown in Table \ref{table:parity}. However, since both strong topological index (\ref{equ:strongTI}) and weak topological indices ({\ref{equ:weakTI}}) contain an even number of L points, the product of $\delta_{\Gamma_i}$ always gives a $+$ sign. Therefore, in order to achieve topological non-trivial phases, it is essential to reduce the number of equivalent L points, which can be achieved by a superlattice structure, as discussed in detail below.

\begin{table}[htbp]
  \caption{Parity and irreducible representation table $\xi(\Gamma_{i})$ of occupied states at TRIM $\Gamma _{i}$.    $\delta_{\Gamma_i}$ is the parity product of occupied states at $\Gamma _{i}$. There are eight TRIM, one $\Gamma$ point, 3 equivalent X points, 4 equilvalent L points.}
  \begin{tabular}{p{2.75cm}p{3.5cm}p{2cm}}
    \hline
    \hline
    PbTe        &    $\xi(\Gamma_{i})$   &  $\delta_{\Gamma_i}$    \\
    1$\Gamma$   &   $\Gamma_{6}^{+}\Gamma_{6}^{+}\Gamma_{6}^{-}2\Gamma_{8}^{-}$    &   $-$   \\
    3X          &   $\mathrm{X_{6}^{+} X_{6}^{+} X_{6}^{-} X_{6}^{-} X_{7}^{-}}$            &   $-$   \\
    4L          &   $\mathrm{L_{6}^{-} L_{6}^{+} L_{6}^{+} L_{45}^{+} L_{6}^{+}}$           &   $-$   \\
    $Z_{2}$ index  &   (0;000)  &  \\
    \hline
    SnTe        &    $\xi(\Gamma_{i})$   &  $\delta_{\Gamma_i}$    \\
    1$\Gamma$   &   $\Gamma_{6}^{+}\Gamma_{6}^{+}\Gamma_{6}^{-}2\Gamma_{8}^{-}$    &   $-$   \\
    3X          &   $\mathrm{X_{6}^{+} X_{6}^{+} X_{6}^{-} X_{6}^{-} X_{7}^{-}}$            &   $-$   \\
    4L          &   $\mathrm{L_{6}^{-} L_{6}^{+} L_{6}^{+} L_{45}^{+} L_{6}^{-}}$           &   +   \\
    $Z_{2}$ index  &   (0;000)  &  \\
    \hline
    \hline
  \end{tabular}
  \label{table:parity}
\end{table}

\section{Weak topological insulators in PbTe/SnTe superlattices}

\begin{figure}[h]
\subfigure{
\includegraphics[width=3in]{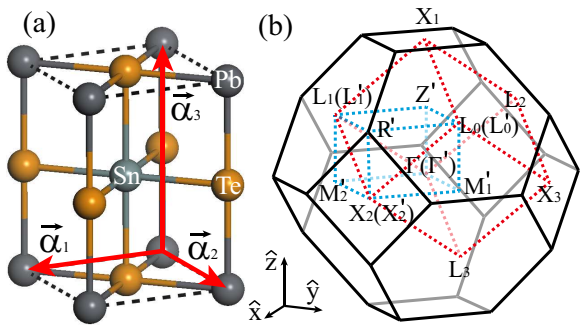}
}
\caption{(a) The primitive cell of $\mathrm{(PbTe)_{1}(SnTe)_{1}}$ [001] superlattice.  (b) Brillouin zone (BZ) and TRIM for PbTe and $\mathrm{(PbTe)_{1}(SnTe)_{1}}$ [001] superlattice: The black lines are the BZ of PbTe primitive cell with the eight TRIM marked by unprimed Greek letters (the red parallelopiped); The vertices of the blue parallelopiped are TRIM marked with primed Greek letters in the BZ of the superlattice.
}
\label{pic:fig1}
\end{figure}

\begin{figure}[h]
\centering
\subfigure{
\includegraphics[width=3.2in]{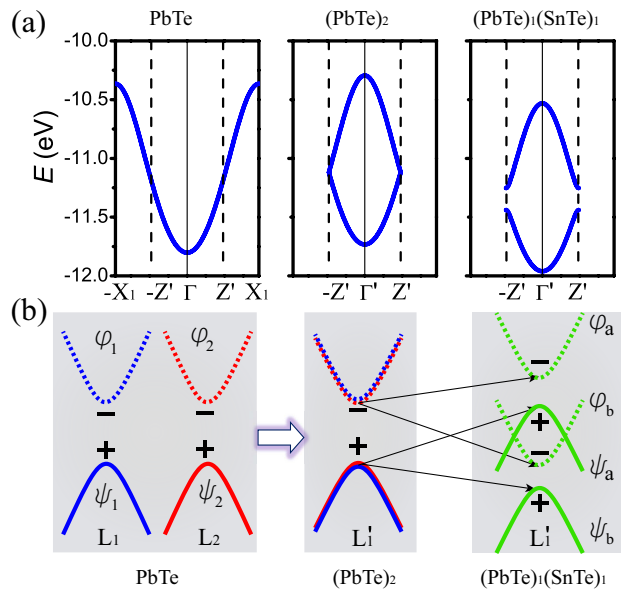}
}
\caption{(a) Band structure evolution from PbTe to $\mathrm{(PbTe)_{1}(SnTe)_{1}}$ superlattice, by taking Te 5s band along the $\Gamma X_1$ as an example. (b) Schematic evolution of energy levels around the Fermi surface, which leads to band inversion. In a superlattice structure, the states from L$_{1}$ and L$_{2}$ are brought together and the degeneracy is removed by doping Sn, resulting in a band inversion. The $+$ or $-$ sign denotes the parity of state at TRIM.}
\label{pic:fig2}
\end{figure}

The band structure of a PbTe/SnTe superlattice can be understood from a simple physical picture in two steps. Let us consider the (PbTe)$_1$(SnTe)$_1$ superlattice along the [001] direction as an example. The first step is to transform from the primitive cell of bulk PbTe to the $\mathrm{(PbTe)_{1}(PbTe)_{1}}$ superlattice cell along the [001] direction. The corresponding primitive lattice vectors are changed from ${\vec{a}_{1}=a(\hat{y} +\hat{z});\vec{a}_{2}=a(\hat{z}+\hat{x});\vec{a}_{3}=a(\hat{x}+\hat{y})}$ of a face-centered cubic lattice, where $a$ is the distance between Pb and nearest Te atom, to ${\vec{\alpha}_{1}=a(\hat{x}-\hat{y});\vec{\alpha}_{2}=a(\hat{x}+\hat{y});\vec{\alpha}_{3}=2a\hat{z}}$ for the superlattice cell, as shown in Fig. \ref{pic:fig1}(a). Since the superlattice cell is twice the primitive cell, the corresponding BZ of the superlattice is folded and becomes half the original one. As shown in Fig. \ref{pic:fig1}, eight TRIM ($\Gamma_i=\Gamma$, 3X, 4L) are transformed to four TRIM $\Gamma_i^{\prime}$ ($i=1,\cdots,4$) in the folded BZ: $\Gamma$ and $\mathrm{X_{1}}$ is projected to a single point $\Gamma^{\prime}_1=\Gamma^{\prime}$, $\mathrm{X_{2}}$ and $\mathrm{X_{3}}$ to $\Gamma^{\prime}_2=\mathrm{X_{2}^{\prime}}$, $\mathrm{L_{0}}$ and L$_{3}$ to $\Gamma^{\prime}_3=\mathrm{L}_{0}^{\prime}$, L$_{1}$ and L$_{2}$ to $\Gamma^{\prime}_4=\mathrm{L}_{1}^{\prime}$. Besides, there will be four new TRIM in the folded BZ, denoted as $\Lambda^{\prime}_{1,2,3,4}=\mathrm{Z}^{\prime},\mathrm{R}^{\prime},\mathrm{M}_{1}^{\prime},\mathrm{M}_{2}^{\prime}$ in Fig. \ref{pic:fig1}(b).
Since the BZ is reduced, band dispersion should also be folded (see Fig. \ref{pic:fig2}(a)).
Consequently, the $\delta_{\Gamma^{\prime}_i}$s in the folded BZ are just the product of the $\delta_{\Gamma_i}$ at the corresponding TRIM in the original BZ, e.g. $\delta_{\Gamma^{\prime}}=\delta_{\Gamma}\delta_{\mathrm{X}_{1}}=1$. The new emerging TRIM $\Lambda_i^{\prime}$ are at the boundary of the folded BZ, so one can combine the wavefunction at $\Lambda_i^{\prime}$ with that at $-\Lambda_i^{\prime}$ to form the bonding and anti-bonding states that are the eigenstates of inversion operation. Since the bonding and anti-bonding states have opposite parities, the $\delta_{\Lambda^{\prime}_i}$s at these new TRIM take the value of $-1$. Thus, from $\delta_{\Gamma^{\prime}_i}=1$  ($i=1,2,3,4$) and $\delta_{\Lambda^{\prime}_i}=-1$  ($i=1,2,3,4$), one finds that both strong topological index and weak topological indices remain unchanged, which is expected since the lattice remains the same in this step.

Next, we substitute one Pb atom by one Sn atom in the superlattice cell as shown in Fig. \ref{pic:fig1}(a). This does not break inversion symmetry and thus we can still use Eqs. (\ref{equ:strongTI}) and (\ref{equ:weakTI}) to evaluate topological invariants. Introducing Sn atoms induces the interaction between the states at $\mathrm{L}^{\prime}_{0,1}$ and splits the degeneracy. When the splitting is large enough, band inversion occurs at these momenta.
In Fig. \ref{pic:fig2}(b), we denote the wavefunctions of conduction bands as $\varphi_{1,2}$ (odd parity) and those of valence bands as $\psi_{1,2}$  (even parity) at L$_{1,2}$. After replacing atoms, both $\varphi_{1,2}$ and $\psi_{1,2}$ are no longer the eigen-states of the (PbTe)$_1$(SnTe)$_1$ superlattice and they will hybridize to form new eigenstates $\varphi_{a,b}$ and $\psi_{a,b}$, as shown in Fig. \ref{pic:fig2}(b). Due to level repulsion, the state $\psi_a$ of the valence band maximum will be pushed up while the state $\varphi_b$ of the conduction band minimum will be pushed down. Since $\psi_a$ and $\varphi_b$ have opposite parities, band inversion occurs when $\psi_a$ and $\varphi_b$ change their sequences, as shown in the second step of Fig. \ref{pic:fig2}(b). A similar situation happens at L$_0^{\prime}$. Thanks to reduction in the number of L points from four to two in the first step, the band inversion at $\mathrm{L}_{0,1}^{\prime}$ can lead to a weak TI phase in PbTe/SnTe superlattices. The replacement by Sn atoms will also split the degeneracy at TRIM $\Lambda^{\prime}_i$, as shown in Fig. \ref{pic:fig2}(a). However, since the initial gaps at these TRIM are huge, the splitting will not change any band sequences. It should be emphasized that this mechanism does \emph{not} rely on the inverted band structure of SnTe. Instead, band inversion originates from the strong coupling between the states at equivalent L points due to the folding of the BZ in a superlattice structure.

\begin{figure}[t]
\centering
\subfigure{
\includegraphics[width=3in]{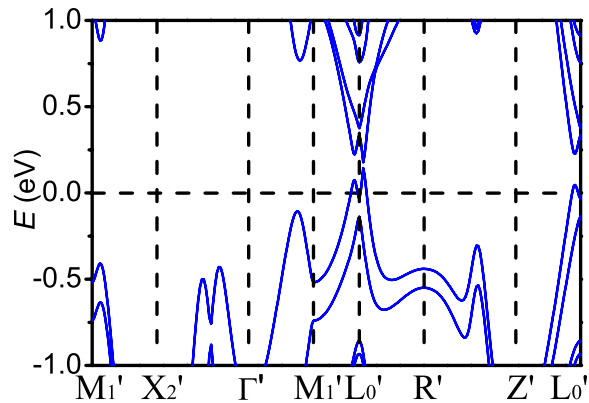}
}
\caption{The band structure of $\mathrm{(PbTe)_{1}(SnTe)_{1}}$ [001] superlattice, with the high symmetry points defined in Figure \ref{pic:fig1}(b). The Mexican-hat shape of dispersion around $L_0'$ indicates a band inversion. The band gap is at the $L_0'-R'$ line.
 }
\label{pic:fig3}
\end{figure}

Therefore, the replacement of Pb atoms by Sn atoms will change the sign of $\delta_{\mathrm{L}^{\prime}_0}$ and $\delta_{\mathrm{L}^{\prime}_1}$ but leave the $\delta_{\Gamma^{\prime}_i}$ and $\delta_{\Lambda^{\prime}_i}$ at other TRIM unchanged. Strong topological index should still be zero since there are always even number times of band inversion. Nevertheless, weak topological indices can be nonzero, so we carry out an {\it ab initio} calculation for the (PbTe)$_1$(SnTe)$_1$ superlattice and the energy dispersion is shown in Fig. \ref{pic:fig3}. A Mexican-hat shape of dispersion appears around L$_0^{\prime}$, indicating the occurrence of band inversion. Furthermore, we check the $\delta_{\Gamma_i}$ at all TRIM $\Gamma^{\prime}_i$ and $\Lambda^{\prime}_i$, as shown in the table \ref{table:parity2}. We find the weak TI indices $\bar{\nu}=(110)$, so the present system can be viewed as a stacking of two dimensional TIs along x direction.  ( Since $\mathrm{\vec{\beta_{1}}=\pi(\hat{x}-\hat{y})/a}$, $\mathrm{\vec{\beta_{2}}=\pi(\hat{x}+\hat{y})/a}$, $\mathrm{\vec{\beta_{3}}=\pi\hat{z}/a}$ are the reciprocal lattice vectors for the (PbTe)$_{1}$(SnTe)$_{1}$ superlattice,  $\vec\beta_{1}+\vec\beta_{2}$  equals  $2\pi\hat{x}/a$). 

\begin{figure}[t]
\centering
\subfigure{
\includegraphics[width=3in]{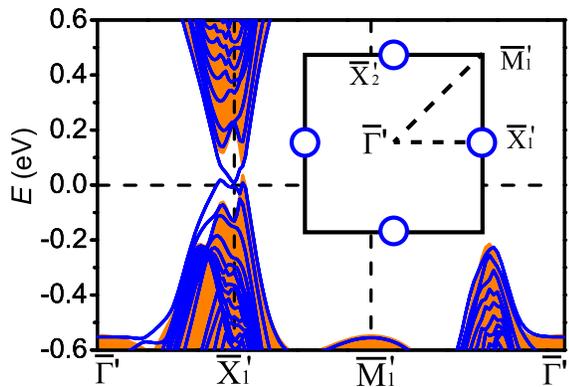}
}

\caption{The energy dispersion of a slab configuration for the $\mathrm{(PbTe)_{1}(SnTe)_{1}}$ [001] superlattice. The shadow indicates the regime of bulk dispersion. A surface state with a Dirac cone appears in the bulk band gap at $\bar{\rm X}'_1$.  The inset shows the surface BZ of the slab.   ${\bar{\Gamma }}^{\prime}$,  $\mathrm{{\bar{{\rm X}}_{1}}^{\prime}}$,        $\mathrm{{\bar{M}_{1}}^{\prime}}$ and $\mathrm{{\bar{{X}}_{2}}^{\prime}}$ are the projections from $\Gamma ^{\prime}$, M$_{1}^{\prime}$, X$_{2}^{\prime}$ and  M$_{2}^{\prime}$ on the (001) surface BZ. L$_{0}^{\prime}$ is projected to ${\mathrm{\bar{X}_{1}}^{\prime}}$.  ${\mathrm{\bar{X}_{1}}^{\prime}}$ and ${\mathrm{\bar{\rm X}_{2}}^{\prime}}$ are equivalent due to the mirror symmetry  with respect to the (100) plane (the plane along the $\bar{\Gamma}'-\bar{\rm M}_1'$ line).}
\label{pic:fig5}
\end{figure}

For a weak TI with $\bar{\nu}=(110)$, surface states with even number of Dirac cones are expected on the surface with Miller index $h \neq \bar{\nu}(mod2)$. Thus, we consider a slab configuration of (PbTe)$_1$(SnTe)$_1$ superlattice along [001] direction and directly calculate surface states with the {\it ab initio} method. Indeed, as shown in Fig. \ref{pic:fig5}, surface states are found around $\rm\bar{X}^{\prime}_1$, which is the projection of $\mathrm{L_{0}^{'}}$ at the surface BZ of the superlattice. The Dirac point is located exactly at TRIM $\rm\bar{X}_1^{\prime}$ (The tiny gap of surface states at $\rm\bar{X}_1^{\prime}$ is due to the finite size effect of a slab configuration). According to the mirror symmetry or the four-fold rotation symmetry that relates $\rm\bar{X}_1^{\prime}$ to $\rm\bar{X}_2^{\prime}$, we expect another Dirac point at TRIM $\rm\bar{X}_2^{\prime}$. The degeneracy at $\rm\bar{X}^{\prime}_{1,2}$ is protected by TR symmetry according to Kramers' theorem.
Therefore, our calculation of topological surface states is consistent with the analysis of bulk topological invariants, confirming that the (PbTe)$_1$(SnTe)$_1$ superlattice is a weak TI.
Similar to the case of strong TIs, the backscattering in one Dirac cone is completely suppressed due to the helical nature of spin texture. Since two $\rm\bar{X}^{\prime}$ points are well separated in momentum space, the scattering between two Dirac cones is negligible for impurities with smooth pontentials.

Remarkably, the surface states here are qualitatively different from the surface states of SnTe, which consist of \emph{four} Dirac points at \emph{non-}TRIM and are protected by \emph{mirror} symmetry instead of TR symmetry. For the (PbTe)$_1$(SnTe)$_1$ superlattice, mirror symmetry can also play a role. Actually, there is additional protection of the gapless Dirac points at $\rm\bar{X}_1^{\prime}$ and $\rm\bar{X}_2^{\prime}$ by the mirror symmetry with respect to $(1\bar{1}0)$ plane (the plane along the line $\bar{\Gamma}'-\bar{X}'_1$ in Fig. \ref{pic:fig5} and perpendicular to the surface) and (110) plane (the plane along the line $\bar{\Gamma}'-\bar{X}'_2$ in Fig. \ref{pic:fig5}), respectively. Since there is only one Dirac cone at one mirror plane, the mirror Chern number $C_m$ should be $1$ in the present system , in contrast to $C_m=2$ in bulk SnTe. Therefore, the (PbTe)$_1$(SnTe)$_1$ superlattice can also be regarded as a TCI with mirror Chern number $C_m=$1. When TR symmetry is broken but mirror symmetry is preserved, e.g. with an in-plane magnetic field along the [1$\bar{1}0$] or [110] direction, the gapless nature of Dirac cones should still remain.

3D weak TIs are usually constructed by stacking 2D TIs, such as layered semiconductors discussed in Ref. [\onlinecite{PhysRevLett109116406,2013arXiv13078054T}]. In these cases, if we take the stacking direction as $z$ direction, the corresponding weak topological indices are $(001)$. In contrast, the weak topological indices $(110)$ of our system are different from the growth direction $(001)$ of superlattices. Thus, the underlying mechanism of our system is not the stacking of 2D TIs, but the folding of BZ, as discussed above. In our system, two surface Dirac cones appearing at $\bar{X}'_1$ and $\bar{X}'_2$ of $(001)$ surfaces are related to each other by four-fold rotation symmetry or mirror symmetry. When there is scattering between two Dirac cones, charge density waves can occur at (001) surfaces, giving rise to the half quantum spin Hall effect proposed in Ref. [\onlinecite{Liu2012906}].


\begin{table}[htbp]
  \caption{Parity table $\xi(\Gamma_{i})$ of occupied states at TRIM $\Gamma _{i}$ for the $\mathrm{(PbTe)_{1}(SnTe)_{1}}$ superlattice. $\delta_{\Gamma_i}$ is the parity product of occupied states at $\Gamma _{i}$. There are eight TRIM: $\mathrm{\Gamma ^{\prime},  X_{2}^{\prime}, L_{0}^{\prime}, L_{1}^{\prime}, Z^{\prime}, R^{\prime}, M_{1}^{\prime}, M_{2}^{\prime}}$. L$_{0}^{\prime}$ and L$_{1}^{\prime}$ are equivalent;  M$_{1}^{\prime}$ and M$_{2}^{\prime}$ are equivalent.  }
  \begin{tabular}{p{2.75cm}p{3.75cm}p{1.6cm}}
    \hline
    \hline
    $\mathrm{(PbTe)_{1}(SnTe)_{1}}$        &    $\xi(\Gamma_{i})$   &  $\delta_{\Gamma_i}$    \\
    $\Gamma ^{\prime}$   &   $+ + + + - - - - - -$    &   $+$   \\
    X$_{2}^{\prime}$     &   $+ + + + - - - - - -$    &   $+$   \\
    $\mathrm{L_{0}^{\prime}(L_{1}^{\prime})}$       &  $- - + + + + + + + -$           &   $-$   \\
    Z$^{\prime}$         &   $+ - + - - + + - + -$    &   $-$   \\
    R$^{\prime}$         &   $- + + - + - - + + -$    &   $-$   \\
    $\mathrm{M_{1}^{\prime}(M_{2}^{\prime})}$       &  $+ - + - - + - + + -$           &   $-$   \\
    $Z_{2}$ index  &   (0;110)  &  \\
    \hline
    \hline
  \end{tabular}
  \label{table:parity2}
\end{table}

Similar discussion can also be applied to other (PbTe)$_m$(SnTe)$_{2n-m}$ superlattices ($n$ and $m$ are integers) and it turns out that the weak TI phase is quite robust. The BZ of (PbTe)$_m$(SnTe)$_{2n-m}$ superlattice can be obtained by simply folding the BZ of a (PbTe)$_{1}$(SnTe)$_{1}$ superlattice along the z direction. Two L$^{\prime}$ points in a (PbTe)$_{1}$(SnTe)$_{1}$ superlattice will be still mapped to two separate TRIM in the new BZ. Thus, the machanism for the weak TI phase is still applicable. As shown in Table \ref{table:super}, for a large range of the ratio $x=\frac{m}{2n}$, the system keeps in the weak TI phase with $\bar{\nu}=(110)$ and the corresponding band gaps vary around tens of meV. Thus, a fine tuning of layer numbers of the superlattice is not necessary. Similar superlattices have been fabricated in early experiments. \cite{Springholz2013263,Fujiyasu1984579,PhysRevB303394,ishida1901}  Therefore, PbTe/SnTe superlattices along the [001] direction provide us an experimentally feasible and controllable platform to investigate the exotic phenomena of weak TIs. Moreover, since our basic mechanism is quite general, it is also worthwhile to investigate GeTe/SnTe \cite{PhysRevB88045207} and PbSe/SnSe superlattices\cite{nmat3449}.

\begin{table}[tbp]
  \caption{Gaps at the momenta $\mathrm{L'}$ and bulk gaps for different superlattices along the [001] direction. Here $\mathrm{L'}$ denotes the momenta in the folded BZ of the superlattice that are projected from the L points in the original BZ of a bulk system. }
  \begin{tabular}{p{2.9cm}ll}
    \hline
    \hline
    Composition                     &   Gap at $\mathrm{L'}$ (meV)  &     Bulk Gap (meV)    \\
     $\mathrm{(PbTe)_{1}(SnTe)_{1}}$          &     370.8        &     33.0             \\
     $\mathrm{(PbTe)_{3}(SnTe)_{1}}$      &     144.6        &     31.8             \\
     $\mathrm{(PbTe)_{1}(SnTe)_{3}}$      &     154.6        &     1.0              \\
     $\mathrm{(PbTe)_{5}(SnTe)_{1}}$      &     87.9         &     26.5             \\
     $\mathrm{(PbTe)_{3}(SnTe)_{3}}$      &     96.9         &     27.5             \\
     $\mathrm{(PbTe)_{1}(SnTe)_{5}}$      &     69.0         &     17.6             \\
    \hline
    \hline
  \end{tabular}
  \label{table:super}
\end{table}

\section{Conclusion}
In summary, we propose a series of ${\mathrm{(PbTe)}_{m}\mathrm{(SnTe)}_{2n-m}}$ superlattice systems to realize weak TIs. Due to the BZ folding, we reduce the number of equivalent L points so that weak TI phases can be realized in PbTe/SnTe superlattices, which cannot be achieved in the bulk Pb$_x$Sn$_{1-x}$Te with uniform doping. We notice that the PbTe/SnTe superlattice along the [111] direction has been investigated with the effective Hamiltonian at four equivalent L points.\cite{PhysRevB85205319} In this case, four L points are projected into different momenta in the folded BZ, so that they can be treated separately and the effect of BZ folding is not important. But for the superlattice along the [001] direction, different L points will be mapped to the same momentum in the folded BZ. Therefore, the coupling between different L points cannot be neglected and instead shows a new mechanism to engineer topological phases. This idea can be generalized to search for new topological phases in other systems where band gap occurs at several equivalent momenta.

\begin{acknowledgments}
 We acknowledge support from the Ministry of Science and Technology of China (Grant Nos.  2011CB921901 and 2011CB606405) and the National Natural Science Foundation of China (Grant No. 11074139). LF is supported by the DOE Office of Basic Energy Sciences, Division of Materials Sciences and Engineering under award DE-SC0010526. 
\end{acknowledgments}


\bibliography{weakTI}

\begin{thebibliography}{62}
\expandafter\ifx\csname natexlab\endcsname\relax\def\natexlab#1{#1}\fi
\expandafter\ifx\csname bibnamefont\endcsname\relax
  \def\bibnamefont#1{#1}\fi
\expandafter\ifx\csname bibfnamefont\endcsname\relax
  \def\bibfnamefont#1{#1}\fi
\expandafter\ifx\csname citenamefont\endcsname\relax
  \def\citenamefont#1{#1}\fi
\expandafter\ifx\csname url\endcsname\relax
  \def\url#1{\texttt{#1}}\fi
\expandafter\ifx\csname urlprefix\endcsname\relax\def\urlprefix{URL }\fi
\providecommand{\bibinfo}[2]{#2}
\providecommand{\eprint}[2][]{\url{#2}}

\bibitem[{\citenamefont{Hasan and Kane}(2010)}]{RevModPhys823045}
\bibinfo{author}{\bibfnamefont{M.~Z.} \bibnamefont{Hasan}} \bibnamefont{and}
  \bibinfo{author}{\bibfnamefont{C.~L.} \bibnamefont{Kane}},
  \bibinfo{journal}{Rev. Mod. Phys.} \textbf{\bibinfo{volume}{82}},
  \bibinfo{pages}{3045} (\bibinfo{year}{2010}).

\bibitem[{\citenamefont{Qi and Zhang}(2011)}]{RevModPhys831057}
\bibinfo{author}{\bibfnamefont{X.-L.} \bibnamefont{Qi}} \bibnamefont{and}
  \bibinfo{author}{\bibfnamefont{S.-C.} \bibnamefont{Zhang}},
  \bibinfo{journal}{Rev. Mod. Phys.} \textbf{\bibinfo{volume}{83}},
  \bibinfo{pages}{1057} (\bibinfo{year}{2011}).

\bibitem[{\citenamefont{Moore}(2010)}]{moore2010}
\bibinfo{author}{\bibfnamefont{J.~E.} \bibnamefont{Moore}},
  \bibinfo{journal}{Nature} \textbf{\bibinfo{volume}{464}},
  \bibinfo{pages}{194} (\bibinfo{year}{2010}).

\bibitem[{\citenamefont{Qi and Zhang}(2010)}]{qi2010phystoday}
\bibinfo{author}{\bibfnamefont{X.~L.} \bibnamefont{Qi}} \bibnamefont{and}
  \bibinfo{author}{\bibfnamefont{S.~C.} \bibnamefont{Zhang}},
  \bibinfo{journal}{Phys. Today} \textbf{\bibinfo{volume}{63}},
  \bibinfo{pages}{33} (\bibinfo{year}{2010}).

\bibitem[{\citenamefont{Kane and Mele}(2005)}]{PhysRevLett95226801}
\bibinfo{author}{\bibfnamefont{C.~L.} \bibnamefont{Kane}} \bibnamefont{and}
  \bibinfo{author}{\bibfnamefont{E.~J.} \bibnamefont{Mele}},
  \bibinfo{journal}{Phys. Rev. Lett.} \textbf{\bibinfo{volume}{95}},
  \bibinfo{pages}{226801} (\bibinfo{year}{2005}).

\bibitem[{\citenamefont{Bernevig and Zhang}(2006)}]{PhysRevLett96106802}
\bibinfo{author}{\bibfnamefont{B.~A.} \bibnamefont{Bernevig}} \bibnamefont{and}
  \bibinfo{author}{\bibfnamefont{S.-C.} \bibnamefont{Zhang}},
  \bibinfo{journal}{Phys. Rev. Lett.} \textbf{\bibinfo{volume}{96}},
  \bibinfo{pages}{106802} (\bibinfo{year}{2006}).

\bibitem[{\citenamefont{Bernevig et~al.}(2006)\citenamefont{Bernevig, Hughes,
  and Zhang}}]{Bernevig15122006}
\bibinfo{author}{\bibfnamefont{B.~A.} \bibnamefont{Bernevig}},
  \bibinfo{author}{\bibfnamefont{T.~L.} \bibnamefont{Hughes}},
  \bibnamefont{and} \bibinfo{author}{\bibfnamefont{S.-C.} \bibnamefont{Zhang}},
  \bibinfo{journal}{Science} \textbf{\bibinfo{volume}{314}},
  \bibinfo{pages}{1757} (\bibinfo{year}{2006}).

\bibitem[{\citenamefont{König et~al.}(2007)\citenamefont{König, Wiedmann,
  Brüne, Roth, Buhmann, Molenkamp, Qi, and Zhang}}]{König02112007}
\bibinfo{author}{\bibfnamefont{M.}~\bibnamefont{König}},
  \bibinfo{author}{\bibfnamefont{S.}~\bibnamefont{Wiedmann}},
  \bibinfo{author}{\bibfnamefont{C.}~\bibnamefont{Brüne}},
  \bibinfo{author}{\bibfnamefont{A.}~\bibnamefont{Roth}},
  \bibinfo{author}{\bibfnamefont{H.}~\bibnamefont{Buhmann}},
  \bibinfo{author}{\bibfnamefont{L.~W.} \bibnamefont{Molenkamp}},
  \bibinfo{author}{\bibfnamefont{X.-L.} \bibnamefont{Qi}}, \bibnamefont{and}
  \bibinfo{author}{\bibfnamefont{S.-C.} \bibnamefont{Zhang}},
  \bibinfo{journal}{Science} \textbf{\bibinfo{volume}{318}},
  \bibinfo{pages}{766} (\bibinfo{year}{2007}).

\bibitem[{\citenamefont{Knez et~al.}(2011)\citenamefont{Knez, Du, and
  Sullivan}}]{PhysRevLett107136603}
\bibinfo{author}{\bibfnamefont{I.}~\bibnamefont{Knez}},
  \bibinfo{author}{\bibfnamefont{R.-R.} \bibnamefont{Du}}, \bibnamefont{and}
  \bibinfo{author}{\bibfnamefont{G.}~\bibnamefont{Sullivan}},
  \bibinfo{journal}{Phys. Rev. Lett.} \textbf{\bibinfo{volume}{107}},
  \bibinfo{pages}{136603} (\bibinfo{year}{2011}).

\bibitem[{\citenamefont{Knez et~al.}(2010)\citenamefont{Knez, Du, and
  Sullivan}}]{PhysRevB81201301}
\bibinfo{author}{\bibfnamefont{I.}~\bibnamefont{Knez}},
  \bibinfo{author}{\bibfnamefont{R.~R.} \bibnamefont{Du}}, \bibnamefont{and}
  \bibinfo{author}{\bibfnamefont{G.}~\bibnamefont{Sullivan}},
  \bibinfo{journal}{Phys. Rev. B} \textbf{\bibinfo{volume}{81}},
  \bibinfo{pages}{201301} (\bibinfo{year}{2010}).

\bibitem[{\citenamefont{{Du} et~al.}(2013)\citenamefont{{Du}, {Knez},
  {Sullivan}, and {Du}}}]{2013arXiv13061925D}
\bibinfo{author}{\bibfnamefont{L.}~\bibnamefont{{Du}}},
  \bibinfo{author}{\bibfnamefont{I.}~\bibnamefont{{Knez}}},
  \bibinfo{author}{\bibfnamefont{G.}~\bibnamefont{{Sullivan}}},
  \bibnamefont{and} \bibinfo{author}{\bibfnamefont{R.-R.} \bibnamefont{{Du}}},
  \bibinfo{journal}{ArXiv e-prints}  (\bibinfo{year}{2013}),
  \eprint{1306.1925}.

\bibitem[{\citenamefont{Fu et~al.}(2007)\citenamefont{Fu, Kane, and
  Mele}}]{PhysRevLett98106803}
\bibinfo{author}{\bibfnamefont{L.}~\bibnamefont{Fu}},
  \bibinfo{author}{\bibfnamefont{C.~L.} \bibnamefont{Kane}}, \bibnamefont{and}
  \bibinfo{author}{\bibfnamefont{E.~J.} \bibnamefont{Mele}},
  \bibinfo{journal}{Phys. Rev. Lett.} \textbf{\bibinfo{volume}{98}},
  \bibinfo{pages}{106803} (\bibinfo{year}{2007}).

\bibitem[{\citenamefont{Fu and Kane}(2007)}]{PhysRevB76045302}
\bibinfo{author}{\bibfnamefont{L.}~\bibnamefont{Fu}} \bibnamefont{and}
  \bibinfo{author}{\bibfnamefont{C.~L.} \bibnamefont{Kane}},
  \bibinfo{journal}{Phys. Rev. B} \textbf{\bibinfo{volume}{76}},
  \bibinfo{pages}{045302} (\bibinfo{year}{2007}).

\bibitem[{\citenamefont{Moore and Balents}(2007)}]{PhysRevB75121306}
\bibinfo{author}{\bibfnamefont{J.~E.} \bibnamefont{Moore}} \bibnamefont{and}
  \bibinfo{author}{\bibfnamefont{L.}~\bibnamefont{Balents}},
  \bibinfo{journal}{Phys. Rev. B} \textbf{\bibinfo{volume}{75}},
  \bibinfo{pages}{121306} (\bibinfo{year}{2007}).

\bibitem[{\citenamefont{Roy}(2009{\natexlab{a}})}]{PhysRevB79195322}
\bibinfo{author}{\bibfnamefont{R.}~\bibnamefont{Roy}}, \bibinfo{journal}{Phys.
  Rev. B} \textbf{\bibinfo{volume}{79}}, \bibinfo{pages}{195322}
  (\bibinfo{year}{2009}{\natexlab{a}}).

\bibitem[{\citenamefont{Zhang et~al.}(2009{\natexlab{a}})\citenamefont{Zhang,
  Liu, Qi, Deng, Dai, Zhang, and Fang}}]{PhysRevB80085307}
\bibinfo{author}{\bibfnamefont{H.-J.} \bibnamefont{Zhang}},
  \bibinfo{author}{\bibfnamefont{C.-X.} \bibnamefont{Liu}},
  \bibinfo{author}{\bibfnamefont{X.-L.} \bibnamefont{Qi}},
  \bibinfo{author}{\bibfnamefont{X.-Y.} \bibnamefont{Deng}},
  \bibinfo{author}{\bibfnamefont{X.}~\bibnamefont{Dai}},
  \bibinfo{author}{\bibfnamefont{S.-C.} \bibnamefont{Zhang}}, \bibnamefont{and}
  \bibinfo{author}{\bibfnamefont{Z.}~\bibnamefont{Fang}},
  \bibinfo{journal}{Phys. Rev. B} \textbf{\bibinfo{volume}{80}},
  \bibinfo{pages}{085307} (\bibinfo{year}{2009}{\natexlab{a}}).

\bibitem[{\citenamefont{Hsieh et~al.}(2008)\citenamefont{Hsieh, Qian, Wray,
  Xia, Hor, Cava, and Hasan}}]{nature06843}
\bibinfo{author}{\bibfnamefont{D.}~\bibnamefont{Hsieh}},
  \bibinfo{author}{\bibfnamefont{D.}~\bibnamefont{Qian}},
  \bibinfo{author}{\bibfnamefont{L.}~\bibnamefont{Wray}},
  \bibinfo{author}{\bibfnamefont{Y.}~\bibnamefont{Xia}},
  \bibinfo{author}{\bibfnamefont{Y.~S.} \bibnamefont{Hor}},
  \bibinfo{author}{\bibfnamefont{R.~J.} \bibnamefont{Cava}}, \bibnamefont{and}
  \bibinfo{author}{\bibfnamefont{M.~Z.} \bibnamefont{Hasan}},
  \bibinfo{journal}{Nature} \textbf{\bibinfo{volume}{452}},
  \bibinfo{pages}{970974} (\bibinfo{year}{2008}).

\bibitem[{\citenamefont{Zhang et~al.}(2009{\natexlab{b}})\citenamefont{Zhang,
  Liu, Qi, Dai, Fang, and Zhang}}]{Nphys101038}
\bibinfo{author}{\bibfnamefont{H.-J.} \bibnamefont{Zhang}},
  \bibinfo{author}{\bibfnamefont{C.-X.} \bibnamefont{Liu}},
  \bibinfo{author}{\bibfnamefont{X.-L.} \bibnamefont{Qi}},
  \bibinfo{author}{\bibfnamefont{X.}~\bibnamefont{Dai}},
  \bibinfo{author}{\bibfnamefont{Z.}~\bibnamefont{Fang}}, \bibnamefont{and}
  \bibinfo{author}{\bibfnamefont{S.-C.} \bibnamefont{Zhang}},
  \bibinfo{journal}{Nat Phys.} \textbf{\bibinfo{volume}{5}},
  \bibinfo{pages}{438} (\bibinfo{year}{2009}{\natexlab{b}}).

\bibitem[{\citenamefont{Chen et~al.}(2009)\citenamefont{Chen, Analytis, Chu,
  Liu, Mo, Qi, Zhang, Lu, Dai, Fang et~al.}}]{Chen10072009}
\bibinfo{author}{\bibfnamefont{Y.~L.} \bibnamefont{Chen}},
  \bibinfo{author}{\bibfnamefont{J.~G.} \bibnamefont{Analytis}},
  \bibinfo{author}{\bibfnamefont{J.-H.} \bibnamefont{Chu}},
  \bibinfo{author}{\bibfnamefont{Z.~K.} \bibnamefont{Liu}},
  \bibinfo{author}{\bibfnamefont{S.-K.} \bibnamefont{Mo}},
  \bibinfo{author}{\bibfnamefont{X.~L.} \bibnamefont{Qi}},
  \bibinfo{author}{\bibfnamefont{H.~J.} \bibnamefont{Zhang}},
  \bibinfo{author}{\bibfnamefont{D.~H.} \bibnamefont{Lu}},
  \bibinfo{author}{\bibfnamefont{X.}~\bibnamefont{Dai}},
  \bibinfo{author}{\bibfnamefont{Z.}~\bibnamefont{Fang}}, \bibnamefont{et~al.},
  \textbf{\bibinfo{volume}{325}}, \bibinfo{pages}{178} (\bibinfo{year}{2009}).

\bibitem[{\citenamefont{Hsieh et~al.}(2009)\citenamefont{Hsieh, Xia, Qian,
  Wray, Meier, Dil, Osterwalder, Patthey, Fedorov, Lin
  et~al.}}]{PhysRevLett103146401}
\bibinfo{author}{\bibfnamefont{D.}~\bibnamefont{Hsieh}},
  \bibinfo{author}{\bibfnamefont{Y.}~\bibnamefont{Xia}},
  \bibinfo{author}{\bibfnamefont{D.}~\bibnamefont{Qian}},
  \bibinfo{author}{\bibfnamefont{L.}~\bibnamefont{Wray}},
  \bibinfo{author}{\bibfnamefont{F.}~\bibnamefont{Meier}},
  \bibinfo{author}{\bibfnamefont{J.~H.} \bibnamefont{Dil}},
  \bibinfo{author}{\bibfnamefont{J.}~\bibnamefont{Osterwalder}},
  \bibinfo{author}{\bibfnamefont{L.}~\bibnamefont{Patthey}},
  \bibinfo{author}{\bibfnamefont{A.~V.} \bibnamefont{Fedorov}},
  \bibinfo{author}{\bibfnamefont{H.}~\bibnamefont{Lin}}, \bibnamefont{et~al.},
  \bibinfo{journal}{Phys. Rev. Lett.} \textbf{\bibinfo{volume}{103}},
  \bibinfo{pages}{146401} (\bibinfo{year}{2009}).

\bibitem[{\citenamefont{Chen et~al.}(2010)\citenamefont{Chen, Liu, Analytis,
  Chu, Zhang, Yan, Mo, Moore, Lu, Fisher et~al.}}]{PhysRevLett105266401}
\bibinfo{author}{\bibfnamefont{Y.~L.} \bibnamefont{Chen}},
  \bibinfo{author}{\bibfnamefont{Z.~K.} \bibnamefont{Liu}},
  \bibinfo{author}{\bibfnamefont{J.~G.} \bibnamefont{Analytis}},
  \bibinfo{author}{\bibfnamefont{J.-H.} \bibnamefont{Chu}},
  \bibinfo{author}{\bibfnamefont{H.~J.} \bibnamefont{Zhang}},
  \bibinfo{author}{\bibfnamefont{B.~H.} \bibnamefont{Yan}},
  \bibinfo{author}{\bibfnamefont{S.-K.} \bibnamefont{Mo}},
  \bibinfo{author}{\bibfnamefont{R.~G.} \bibnamefont{Moore}},
  \bibinfo{author}{\bibfnamefont{D.~H.} \bibnamefont{Lu}},
  \bibinfo{author}{\bibfnamefont{I.~R.} \bibnamefont{Fisher}},
  \bibnamefont{et~al.}, \bibinfo{journal}{Phys. Rev. Lett.}
  \textbf{\bibinfo{volume}{105}}, \bibinfo{pages}{266401}
  (\bibinfo{year}{2010}).

\bibitem[{\citenamefont{Sato et~al.}(2010)\citenamefont{Sato, Segawa, Guo,
  Sugawara, Souma, Takahashi, and Ando}}]{PhysRevLett105136802}
\bibinfo{author}{\bibfnamefont{T.}~\bibnamefont{Sato}},
  \bibinfo{author}{\bibfnamefont{K.}~\bibnamefont{Segawa}},
  \bibinfo{author}{\bibfnamefont{H.}~\bibnamefont{Guo}},
  \bibinfo{author}{\bibfnamefont{K.}~\bibnamefont{Sugawara}},
  \bibinfo{author}{\bibfnamefont{S.}~\bibnamefont{Souma}},
  \bibinfo{author}{\bibfnamefont{T.}~\bibnamefont{Takahashi}},
  \bibnamefont{and} \bibinfo{author}{\bibfnamefont{Y.}~\bibnamefont{Ando}},
  \bibinfo{journal}{Phys. Rev. Lett.} \textbf{\bibinfo{volume}{105}},
  \bibinfo{pages}{136802} (\bibinfo{year}{2010}).

\bibitem[{\citenamefont{{Xu} et~al.}(2010)\citenamefont{{Xu}, {Wray}, {Xia},
  {Shankar}, {Petersen}, {Fedorov}, {Lin}, {Bansil}, {Hor}, {Grauer}
  et~al.}}]{2010arXiv10075111X}
\bibinfo{author}{\bibfnamefont{S.-Y.} \bibnamefont{{Xu}}},
  \bibinfo{author}{\bibfnamefont{L.~A.} \bibnamefont{{Wray}}},
  \bibinfo{author}{\bibfnamefont{Y.}~\bibnamefont{{Xia}}},
  \bibinfo{author}{\bibfnamefont{R.}~\bibnamefont{{Shankar}}},
  \bibinfo{author}{\bibfnamefont{A.}~\bibnamefont{{Petersen}}},
  \bibinfo{author}{\bibfnamefont{A.}~\bibnamefont{{Fedorov}}},
  \bibinfo{author}{\bibfnamefont{H.}~\bibnamefont{{Lin}}},
  \bibinfo{author}{\bibfnamefont{A.}~\bibnamefont{{Bansil}}},
  \bibinfo{author}{\bibfnamefont{Y.~S.} \bibnamefont{{Hor}}},
  \bibinfo{author}{\bibfnamefont{D.}~\bibnamefont{{Grauer}}},
  \bibnamefont{et~al.}, \bibinfo{journal}{ArXiv e-prints}
  (\bibinfo{year}{2010}), \eprint{1007.5111}.

\bibitem[{\citenamefont{Jin et~al.}(2011)\citenamefont{Jin, Song, Freeman, and
  Kanatzidis}}]{PhysRevB83041202}
\bibinfo{author}{\bibfnamefont{H.}~\bibnamefont{Jin}},
  \bibinfo{author}{\bibfnamefont{J.-H.} \bibnamefont{Song}},
  \bibinfo{author}{\bibfnamefont{A.~J.} \bibnamefont{Freeman}},
  \bibnamefont{and} \bibinfo{author}{\bibfnamefont{M.~G.}
  \bibnamefont{Kanatzidis}}, \bibinfo{journal}{Phys. Rev. B}
  \textbf{\bibinfo{volume}{83}}, \bibinfo{pages}{041202}
  (\bibinfo{year}{2011}).

\bibitem[{\citenamefont{Br\"une et~al.}(2011)\citenamefont{Br\"une, Liu, Novik,
  Hankiewicz, Buhmann, Chen, Qi, Shen, Zhang, and
  Molenkamp}}]{PhysRevLett106126803}
\bibinfo{author}{\bibfnamefont{C.}~\bibnamefont{Br\"une}},
  \bibinfo{author}{\bibfnamefont{C.~X.} \bibnamefont{Liu}},
  \bibinfo{author}{\bibfnamefont{E.~G.} \bibnamefont{Novik}},
  \bibinfo{author}{\bibfnamefont{E.~M.} \bibnamefont{Hankiewicz}},
  \bibinfo{author}{\bibfnamefont{H.}~\bibnamefont{Buhmann}},
  \bibinfo{author}{\bibfnamefont{Y.~L.} \bibnamefont{Chen}},
  \bibinfo{author}{\bibfnamefont{X.~L.} \bibnamefont{Qi}},
  \bibinfo{author}{\bibfnamefont{Z.~X.} \bibnamefont{Shen}},
  \bibinfo{author}{\bibfnamefont{S.~C.} \bibnamefont{Zhang}}, \bibnamefont{and}
  \bibinfo{author}{\bibfnamefont{L.~W.} \bibnamefont{Molenkamp}},
  \bibinfo{journal}{Phys. Rev. Lett.} \textbf{\bibinfo{volume}{106}},
  \bibinfo{pages}{126803} (\bibinfo{year}{2011}).

\bibitem[{\citenamefont{Hancock et~al.}(2011)\citenamefont{Hancock, van
  Mechelen, Kuzmenko, van~der Marel, Br\"une, Novik, Astakhov, Buhmann, and
  Molenkamp}}]{PhysRevLett107136803}
\bibinfo{author}{\bibfnamefont{J.~N.} \bibnamefont{Hancock}},
  \bibinfo{author}{\bibfnamefont{J.~L.~M.} \bibnamefont{van Mechelen}},
  \bibinfo{author}{\bibfnamefont{A.~B.} \bibnamefont{Kuzmenko}},
  \bibinfo{author}{\bibfnamefont{D.}~\bibnamefont{van~der Marel}},
  \bibinfo{author}{\bibfnamefont{C.}~\bibnamefont{Br\"une}},
  \bibinfo{author}{\bibfnamefont{E.~G.} \bibnamefont{Novik}},
  \bibinfo{author}{\bibfnamefont{G.~V.} \bibnamefont{Astakhov}},
  \bibinfo{author}{\bibfnamefont{H.}~\bibnamefont{Buhmann}}, \bibnamefont{and}
  \bibinfo{author}{\bibfnamefont{L.~W.} \bibnamefont{Molenkamp}},
  \bibinfo{journal}{Phys. Rev. Lett.} \textbf{\bibinfo{volume}{107}},
  \bibinfo{pages}{136803} (\bibinfo{year}{2011}).

\bibitem[{\citenamefont{Yan et~al.}(2012)\citenamefont{Yan, M\"uchler, and
  Felser}}]{PhysRevLett109116406}
\bibinfo{author}{\bibfnamefont{B.}~\bibnamefont{Yan}},
  \bibinfo{author}{\bibfnamefont{L.}~\bibnamefont{M\"uchler}},
  \bibnamefont{and} \bibinfo{author}{\bibfnamefont{C.}~\bibnamefont{Felser}},
  \bibinfo{journal}{Phys. Rev. Lett.} \textbf{\bibinfo{volume}{109}},
  \bibinfo{pages}{116406} (\bibinfo{year}{2012}).

\bibitem[{\citenamefont{Hou et~al.}(2011)\citenamefont{Hou, Zhang, and
  Wang}}]{PhysRevB84075105}
\bibinfo{author}{\bibfnamefont{J.-M.} \bibnamefont{Hou}},
  \bibinfo{author}{\bibfnamefont{W.-X.} \bibnamefont{Zhang}}, \bibnamefont{and}
  \bibinfo{author}{\bibfnamefont{G.-X.} \bibnamefont{Wang}},
  \bibinfo{journal}{Phys. Rev. B} \textbf{\bibinfo{volume}{84}},
  \bibinfo{pages}{075105} (\bibinfo{year}{2011}).

\bibitem[{\citenamefont{{Tang} et~al.}(2013)\citenamefont{{Tang}, {Yan}, {Cao},
  {Wu}, {Felser}, and {Duan}}}]{2013arXiv13078054T}
\bibinfo{author}{\bibfnamefont{P.}~\bibnamefont{{Tang}}},
  \bibinfo{author}{\bibfnamefont{B.}~\bibnamefont{{Yan}}},
  \bibinfo{author}{\bibfnamefont{W.}~\bibnamefont{{Cao}}},
  \bibinfo{author}{\bibfnamefont{S.-C.} \bibnamefont{{Wu}}},
  \bibinfo{author}{\bibfnamefont{C.}~\bibnamefont{{Felser}}}, \bibnamefont{and}
  \bibinfo{author}{\bibfnamefont{W.}~\bibnamefont{{Duan}}},
  \bibinfo{journal}{ArXiv e-prints}  (\bibinfo{year}{2013}),
  \eprint{1307.8054}.

\bibitem[{\citenamefont{Ran et~al.}(2009)\citenamefont{Ran, Zhang, and
  Vishwanath}}]{ranyingnaturephys}
\bibinfo{author}{\bibfnamefont{Y.}~\bibnamefont{Ran}},
  \bibinfo{author}{\bibfnamefont{Y.}~\bibnamefont{Zhang}}, \bibnamefont{and}
  \bibinfo{author}{\bibfnamefont{A.}~\bibnamefont{Vishwanath}},
  \bibinfo{journal}{Nat Phys} \textbf{\bibinfo{volume}{5}},
  \bibinfo{pages}{298303} (\bibinfo{year}{2009}).

\bibitem[{\citenamefont{Mong et~al.}(2012)\citenamefont{Mong, Bardarson, and
  Moore}}]{PhysRevLett108076804}
\bibinfo{author}{\bibfnamefont{R.~S.~K.} \bibnamefont{Mong}},
  \bibinfo{author}{\bibfnamefont{J.~H.} \bibnamefont{Bardarson}},
  \bibnamefont{and} \bibinfo{author}{\bibfnamefont{J.~E.} \bibnamefont{Moore}},
  \bibinfo{journal}{Phys. Rev. Lett.} \textbf{\bibinfo{volume}{108}},
  \bibinfo{pages}{076804} (\bibinfo{year}{2012}).

\bibitem[{\citenamefont{Liu et~al.}(2012)\citenamefont{Liu, Qi, and
  Zhang}}]{Liu2012906}
\bibinfo{author}{\bibfnamefont{C.-X.} \bibnamefont{Liu}},
  \bibinfo{author}{\bibfnamefont{X.-L.} \bibnamefont{Qi}}, \bibnamefont{and}
  \bibinfo{author}{\bibfnamefont{S.-C.} \bibnamefont{Zhang}},
  \bibinfo{journal}{Physica E: Low-dimensional Systems and Nanostructures}
  \textbf{\bibinfo{volume}{44}}, \bibinfo{pages}{906 } (\bibinfo{year}{2012}).

\bibitem[{\citenamefont{Imura et~al.}(2011)\citenamefont{Imura, Takane, and
  Tanaka}}]{PhysRevB84035443}
\bibinfo{author}{\bibfnamefont{K.-I.} \bibnamefont{Imura}},
  \bibinfo{author}{\bibfnamefont{Y.}~\bibnamefont{Takane}}, \bibnamefont{and}
  \bibinfo{author}{\bibfnamefont{A.}~\bibnamefont{Tanaka}},
  \bibinfo{journal}{Phys. Rev. B} \textbf{\bibinfo{volume}{84}},
  \bibinfo{pages}{035443} (\bibinfo{year}{2011}).

\bibitem[{\citenamefont{Imura et~al.}(2012)\citenamefont{Imura, Okamoto,
  Yoshimura, Takane, and Ohtsuki}}]{PhysRevB86245436}
\bibinfo{author}{\bibfnamefont{K.-I.} \bibnamefont{Imura}},
  \bibinfo{author}{\bibfnamefont{M.}~\bibnamefont{Okamoto}},
  \bibinfo{author}{\bibfnamefont{Y.}~\bibnamefont{Yoshimura}},
  \bibinfo{author}{\bibfnamefont{Y.}~\bibnamefont{Takane}}, \bibnamefont{and}
  \bibinfo{author}{\bibfnamefont{T.}~\bibnamefont{Ohtsuki}},
  \bibinfo{journal}{Phys. Rev. B} \textbf{\bibinfo{volume}{86}},
  \bibinfo{pages}{245436} (\bibinfo{year}{2012}).

\bibitem[{\citenamefont{Yoshimura et~al.}(2013)\citenamefont{Yoshimura,
  Matsumoto, Takane, and Imura}}]{PhysRevB88045408}
\bibinfo{author}{\bibfnamefont{Y.}~\bibnamefont{Yoshimura}},
  \bibinfo{author}{\bibfnamefont{A.}~\bibnamefont{Matsumoto}},
  \bibinfo{author}{\bibfnamefont{Y.}~\bibnamefont{Takane}}, \bibnamefont{and}
  \bibinfo{author}{\bibfnamefont{K.-I.} \bibnamefont{Imura}},
  \bibinfo{journal}{Phys. Rev. B} \textbf{\bibinfo{volume}{88}},
  \bibinfo{pages}{045408} (\bibinfo{year}{2013}).

\bibitem[{\citenamefont{Ringel et~al.}(2012)\citenamefont{Ringel, Kraus, and
  Stern}}]{PhysRevB86045102}
\bibinfo{author}{\bibfnamefont{Z.}~\bibnamefont{Ringel}},
  \bibinfo{author}{\bibfnamefont{Y.~E.} \bibnamefont{Kraus}}, \bibnamefont{and}
  \bibinfo{author}{\bibfnamefont{A.}~\bibnamefont{Stern}},
  \bibinfo{journal}{Phys. Rev. B} \textbf{\bibinfo{volume}{86}},
  \bibinfo{pages}{045102} (\bibinfo{year}{2012}).

\bibitem[{\citenamefont{{Fulga} et~al.}(2012)\citenamefont{{Fulga}, {van Heck},
  {Edge}, and {Akhmerov}}}]{2012arXiv12126191F}
\bibinfo{author}{\bibfnamefont{I.~C.} \bibnamefont{{Fulga}}},
  \bibinfo{author}{\bibfnamefont{B.}~\bibnamefont{{van Heck}}},
  \bibinfo{author}{\bibfnamefont{J.~M.} \bibnamefont{{Edge}}},
  \bibnamefont{and} \bibinfo{author}{\bibfnamefont{A.~R.}
  \bibnamefont{{Akhmerov}}}, \bibinfo{journal}{ArXiv e-prints}
  (\bibinfo{year}{2012}), \eprint{1212.6191}.

\bibitem[{\citenamefont{Fu and Kane}(2012)}]{PhysRevLett109246605}
\bibinfo{author}{\bibfnamefont{L.}~\bibnamefont{Fu}} \bibnamefont{and}
  \bibinfo{author}{\bibfnamefont{C.~L.} \bibnamefont{Kane}},
  \bibinfo{journal}{Phys. Rev. Lett.} \textbf{\bibinfo{volume}{109}},
  \bibinfo{pages}{246605} (\bibinfo{year}{2012}).

\bibitem[{\citenamefont{Roy}(2009{\natexlab{b}})}]{PhysRevB79195321}
\bibinfo{author}{\bibfnamefont{R.}~\bibnamefont{Roy}}, \bibinfo{journal}{Phys.
  Rev. B} \textbf{\bibinfo{volume}{79}}, \bibinfo{pages}{195321}
  (\bibinfo{year}{2009}{\natexlab{b}}).

\bibitem[{\citenamefont{Dziawa et~al.}(2012)\citenamefont{Dziawa, Kowalski,
  Dybko, Buczko, Szczerbakow, Szot, tusakowska, Balasubramanian, Wojek,
  Berntsen et~al.}}]{nmat3449}
\bibinfo{author}{\bibfnamefont{P.}~\bibnamefont{Dziawa}},
  \bibinfo{author}{\bibfnamefont{B.~J.} \bibnamefont{Kowalski}},
  \bibinfo{author}{\bibfnamefont{K.}~\bibnamefont{Dybko}},
  \bibinfo{author}{\bibfnamefont{R.}~\bibnamefont{Buczko}},
  \bibinfo{author}{\bibfnamefont{A.}~\bibnamefont{Szczerbakow}},
  \bibinfo{author}{\bibfnamefont{M.}~\bibnamefont{Szot}},
  \bibinfo{author}{\bibfnamefont{E.}~\bibnamefont{tusakowska}},
  \bibinfo{author}{\bibfnamefont{T.}~\bibnamefont{Balasubramanian}},
  \bibinfo{author}{\bibfnamefont{B.~M.} \bibnamefont{Wojek}},
  \bibinfo{author}{\bibfnamefont{M.~H.} \bibnamefont{Berntsen}},
  \bibnamefont{et~al.}, \bibinfo{journal}{Nat Mater}
  \textbf{\bibinfo{volume}{11}}, \bibinfo{pages}{10231027}
  (\bibinfo{year}{2012}).

\bibitem[{\citenamefont{Sato et~al.}(2013)\citenamefont{Sato, Tanaka, Nakayama,
  Souma, Takahashi, Sasaki, Ren, Taskin, Segawa, and
  Ando}}]{PhysRevLett110206804}
\bibinfo{author}{\bibfnamefont{T.}~\bibnamefont{Sato}},
  \bibinfo{author}{\bibfnamefont{Y.}~\bibnamefont{Tanaka}},
  \bibinfo{author}{\bibfnamefont{K.}~\bibnamefont{Nakayama}},
  \bibinfo{author}{\bibfnamefont{S.}~\bibnamefont{Souma}},
  \bibinfo{author}{\bibfnamefont{T.}~\bibnamefont{Takahashi}},
  \bibinfo{author}{\bibfnamefont{S.}~\bibnamefont{Sasaki}},
  \bibinfo{author}{\bibfnamefont{Z.}~\bibnamefont{Ren}},
  \bibinfo{author}{\bibfnamefont{A.~A.} \bibnamefont{Taskin}},
  \bibinfo{author}{\bibfnamefont{K.}~\bibnamefont{Segawa}}, \bibnamefont{and}
  \bibinfo{author}{\bibfnamefont{Y.}~\bibnamefont{Ando}},
  \bibinfo{journal}{Phys. Rev. Lett.} \textbf{\bibinfo{volume}{110}},
  \bibinfo{pages}{206804} (\bibinfo{year}{2013}).

\bibitem[{\citenamefont{Wei and Zunger}(1997)}]{PhysRevB5513605}
\bibinfo{author}{\bibfnamefont{S.-H.} \bibnamefont{Wei}} \bibnamefont{and}
  \bibinfo{author}{\bibfnamefont{A.}~\bibnamefont{Zunger}},
  \bibinfo{journal}{Phys. Rev. B} \textbf{\bibinfo{volume}{55}},
  \bibinfo{pages}{13605} (\bibinfo{year}{1997}).

\bibitem[{\citenamefont{Hummer et~al.}(2007)\citenamefont{Hummer, Gr\"uneis,
  and Kresse}}]{PhysRevB75195211}
\bibinfo{author}{\bibfnamefont{K.}~\bibnamefont{Hummer}},
  \bibinfo{author}{\bibfnamefont{A.}~\bibnamefont{Gr\"uneis}},
  \bibnamefont{and} \bibinfo{author}{\bibfnamefont{G.}~\bibnamefont{Kresse}},
  \bibinfo{journal}{Phys. Rev. B} \textbf{\bibinfo{volume}{75}},
  \bibinfo{pages}{195211} (\bibinfo{year}{2007}).

\bibitem[{\citenamefont{Svane et~al.}(2010)\citenamefont{Svane, Christensen,
  Cardona, Chantis, van Schilfgaarde, and Kotani}}]{PhysRevB81245120}
\bibinfo{author}{\bibfnamefont{A.}~\bibnamefont{Svane}},
  \bibinfo{author}{\bibfnamefont{N.~E.} \bibnamefont{Christensen}},
  \bibinfo{author}{\bibfnamefont{M.}~\bibnamefont{Cardona}},
  \bibinfo{author}{\bibfnamefont{A.~N.} \bibnamefont{Chantis}},
  \bibinfo{author}{\bibfnamefont{M.}~\bibnamefont{van Schilfgaarde}},
  \bibnamefont{and} \bibinfo{author}{\bibfnamefont{T.}~\bibnamefont{Kotani}},
  \bibinfo{journal}{Phys. Rev. B} \textbf{\bibinfo{volume}{81}},
  \bibinfo{pages}{245120} (\bibinfo{year}{2010}).

\bibitem[{\citenamefont{Hsieh Timothy H.and~Lin
  et~al.}(2012)\citenamefont{Hsieh Timothy H.and~Lin, Liu, Duan, Arun, and
  Fu}}]{ncomms101038}
\bibinfo{author}{\bibfnamefont{H.}~\bibnamefont{Hsieh Timothy H.and~Lin}},
  \bibinfo{author}{\bibfnamefont{J.}~\bibnamefont{Liu}},
  \bibinfo{author}{\bibfnamefont{W.}~\bibnamefont{Duan}},
  \bibinfo{author}{\bibfnamefont{B.}~\bibnamefont{Arun}}, \bibnamefont{and}
  \bibinfo{author}{\bibfnamefont{L.}~\bibnamefont{Fu}}, \bibinfo{journal}{Nat
  Commun} \textbf{\bibinfo{volume}{3}} (\bibinfo{year}{2012}).

\bibitem[{\citenamefont{Fu}(2011)}]{PhysRevLett106106802}
\bibinfo{author}{\bibfnamefont{L.}~\bibnamefont{Fu}}, \bibinfo{journal}{Phys.
  Rev. Lett.} \textbf{\bibinfo{volume}{106}}, \bibinfo{pages}{106802}
  (\bibinfo{year}{2011}).

\bibitem[{\citenamefont{Fang et~al.}(2012)\citenamefont{Fang, Gilbert, and
  Bernevig}}]{fang2012a}
\bibinfo{author}{\bibfnamefont{C.}~\bibnamefont{Fang}},
  \bibinfo{author}{\bibfnamefont{M.~J.} \bibnamefont{Gilbert}},
  \bibnamefont{and} \bibinfo{author}{\bibfnamefont{B.~A.}
  \bibnamefont{Bernevig}}, \bibinfo{journal}{Physical Review B}
  \textbf{\bibinfo{volume}{86}}, \bibinfo{pages}{115112}
  (\bibinfo{year}{2012}).

\bibitem[{\citenamefont{Slager et~al.}(2012)\citenamefont{Slager, Mesaros,
  Juri{\v{c}}i{\'c}, and Zaanen}}]{slager2012}
\bibinfo{author}{\bibfnamefont{R.-J.} \bibnamefont{Slager}},
  \bibinfo{author}{\bibfnamefont{A.}~\bibnamefont{Mesaros}},
  \bibinfo{author}{\bibfnamefont{V.}~\bibnamefont{Juri{\v{c}}i{\'c}}},
  \bibnamefont{and} \bibinfo{author}{\bibfnamefont{J.}~\bibnamefont{Zaanen}},
  \bibinfo{journal}{Nature Physics} \textbf{\bibinfo{volume}{9}},
  \bibinfo{pages}{98} (\bibinfo{year}{2012}).

\bibitem[{\citenamefont{Liu et~al.}(2013)\citenamefont{Liu, Duan, and Fu}}]{jw}
\bibinfo{author}{\bibfnamefont{J.}~\bibnamefont{Liu}},
  \bibinfo{author}{\bibfnamefont{W.}~\bibnamefont{Duan}}, \bibnamefont{and}
  \bibinfo{author}{\bibfnamefont{L.}~\bibnamefont{Fu}},
  \bibinfo{journal}{arXiv:1304.0430}  (\bibinfo{year}{2013}).

\bibitem[{\citenamefont{Tanaka et~al.}(2012)\citenamefont{Tanaka, Ren, Sato,
  Nakayama, Souma, Takahashi, Segawa, and Ando}}]{nphys2442}
\bibinfo{author}{\bibfnamefont{Y.}~\bibnamefont{Tanaka}},
  \bibinfo{author}{\bibfnamefont{Z.}~\bibnamefont{Ren}},
  \bibinfo{author}{\bibfnamefont{T.}~\bibnamefont{Sato}},
  \bibinfo{author}{\bibfnamefont{K.}~\bibnamefont{Nakayama}},
  \bibinfo{author}{\bibfnamefont{S.}~\bibnamefont{Souma}},
  \bibinfo{author}{\bibfnamefont{T.}~\bibnamefont{Takahashi}},
  \bibinfo{author}{\bibfnamefont{K.}~\bibnamefont{Segawa}}, \bibnamefont{and}
  \bibinfo{author}{\bibfnamefont{Y.}~\bibnamefont{Ando}}, \bibinfo{journal}{Nat
  Phys} \textbf{\bibinfo{volume}{8}} (\bibinfo{year}{2012}).

\bibitem[{\citenamefont{Xu et~al.}(2012)\citenamefont{Xu, Liu, Alidoust,
  Neupane, Qian, Belopolski, Denlinger, Wang, Lin, Wray et~al.}}]{xu2012a}
\bibinfo{author}{\bibfnamefont{S.-Y.} \bibnamefont{Xu}},
  \bibinfo{author}{\bibfnamefont{C.}~\bibnamefont{Liu}},
  \bibinfo{author}{\bibfnamefont{N.}~\bibnamefont{Alidoust}},
  \bibinfo{author}{\bibfnamefont{M.}~\bibnamefont{Neupane}},
  \bibinfo{author}{\bibfnamefont{D.}~\bibnamefont{Qian}},
  \bibinfo{author}{\bibfnamefont{I.}~\bibnamefont{Belopolski}},
  \bibinfo{author}{\bibfnamefont{J.}~\bibnamefont{Denlinger}},
  \bibinfo{author}{\bibfnamefont{Y.}~\bibnamefont{Wang}},
  \bibinfo{author}{\bibfnamefont{H.}~\bibnamefont{Lin}},
  \bibinfo{author}{\bibfnamefont{L.}~\bibnamefont{Wray}}, \bibnamefont{et~al.},
  \bibinfo{journal}{Nature communications} \textbf{\bibinfo{volume}{3}},
  \bibinfo{pages}{1192} (\bibinfo{year}{2012}).

\bibitem[{\citenamefont{Lin et~al.}(2010)\citenamefont{Lin, Markiewicz, Wray,
  Fu, Hasan, and Bansil}}]{PhysRevLett105036404}
\bibinfo{author}{\bibfnamefont{H.}~\bibnamefont{Lin}},
  \bibinfo{author}{\bibfnamefont{R.~S.} \bibnamefont{Markiewicz}},
  \bibinfo{author}{\bibfnamefont{L.~A.} \bibnamefont{Wray}},
  \bibinfo{author}{\bibfnamefont{L.}~\bibnamefont{Fu}},
  \bibinfo{author}{\bibfnamefont{M.~Z.} \bibnamefont{Hasan}}, \bibnamefont{and}
  \bibinfo{author}{\bibfnamefont{A.}~\bibnamefont{Bansil}},
  \bibinfo{journal}{Phys. Rev. Lett.} \textbf{\bibinfo{volume}{105}},
  \bibinfo{pages}{036404} (\bibinfo{year}{2010}).

\bibitem[{\citenamefont{Buczko and Cywi\ifmmode~\acute{n}\else
  \'{n}\fi{}ski}(2012)}]{PhysRevB85205319}
\bibinfo{author}{\bibfnamefont{R.}~\bibnamefont{Buczko}} \bibnamefont{and}
  \bibinfo{author}{\bibfnamefont{L.}~\bibnamefont{Cywi\ifmmode~\acute{n}\else
  \'{n}\fi{}ski}}, \bibinfo{journal}{Phys. Rev. B}
  \textbf{\bibinfo{volume}{85}}, \bibinfo{pages}{205319}
  (\bibinfo{year}{2012}).

\bibitem[{\citenamefont{Perdew et~al.}(1996)\citenamefont{Perdew, Burke, and
  Ernzerhof}}]{PhysRevLett773865}
\bibinfo{author}{\bibfnamefont{J.~P.} \bibnamefont{Perdew}},
  \bibinfo{author}{\bibfnamefont{K.}~\bibnamefont{Burke}}, \bibnamefont{and}
  \bibinfo{author}{\bibfnamefont{M.}~\bibnamefont{Ernzerhof}},
  \bibinfo{journal}{Phys. Rev. Lett.} \textbf{\bibinfo{volume}{77}},
  \bibinfo{pages}{3865} (\bibinfo{year}{1996}).

\bibitem[{\citenamefont{Bl\"ochl}(1994)}]{PhysRevB5017953}
\bibinfo{author}{\bibfnamefont{P.~E.} \bibnamefont{Bl\"ochl}},
  \bibinfo{journal}{Phys. Rev. B} \textbf{\bibinfo{volume}{50}},
  \bibinfo{pages}{17953} (\bibinfo{year}{1994}).

\bibitem[{\citenamefont{Kresse and Furthm\"uller}(1996)}]{PhysRevB5411169}
\bibinfo{author}{\bibfnamefont{G.}~\bibnamefont{Kresse}} \bibnamefont{and}
  \bibinfo{author}{\bibfnamefont{J.}~\bibnamefont{Furthm\"uller}},
  \bibinfo{journal}{Phys. Rev. B} \textbf{\bibinfo{volume}{54}},
  \bibinfo{pages}{11169} (\bibinfo{year}{1996}).

\bibitem[{\citenamefont{Fu and Kane}(2006)}]{PhysRevB74195312}
\bibinfo{author}{\bibfnamefont{L.}~\bibnamefont{Fu}} \bibnamefont{and}
  \bibinfo{author}{\bibfnamefont{C.~L.} \bibnamefont{Kane}},
  \bibinfo{journal}{Phys. Rev. B} \textbf{\bibinfo{volume}{74}},
  \bibinfo{pages}{195312} (\bibinfo{year}{2006}).

\bibitem[{\citenamefont{Springholz}(2013)}]{Springholz2013263}
\bibinfo{author}{\bibfnamefont{G.}~\bibnamefont{Springholz}}, in
  \emph{\bibinfo{booktitle}{Molecular Beam Epitaxy}}
  (\bibinfo{publisher}{Elsevier}, \bibinfo{address}{Oxford},
  \bibinfo{year}{2013}), pp. \bibinfo{pages}{263 -- 310}.

\bibitem[{\citenamefont{Fujiyasu et~al.}(1984)\citenamefont{Fujiyasu, Ishida,
  Kuwabara, Shimomura, Takaoka, and Murase}}]{Fujiyasu1984579}
\bibinfo{author}{\bibfnamefont{H.}~\bibnamefont{Fujiyasu}},
  \bibinfo{author}{\bibfnamefont{A.}~\bibnamefont{Ishida}},
  \bibinfo{author}{\bibfnamefont{H.}~\bibnamefont{Kuwabara}},
  \bibinfo{author}{\bibfnamefont{S.}~\bibnamefont{Shimomura}},
  \bibinfo{author}{\bibfnamefont{S.}~\bibnamefont{Takaoka}}, \bibnamefont{and}
  \bibinfo{author}{\bibfnamefont{K.}~\bibnamefont{Murase}},
  \bibinfo{journal}{Surface Science} \textbf{\bibinfo{volume}{142}},
  \bibinfo{pages}{579 } (\bibinfo{year}{1984}).

\bibitem[{\citenamefont{Kriechbaum et~al.}(1984)\citenamefont{Kriechbaum,
  Ambrosch, Fantner, Clemens, and Bauer}}]{PhysRevB303394}
\bibinfo{author}{\bibfnamefont{M.}~\bibnamefont{Kriechbaum}},
  \bibinfo{author}{\bibfnamefont{K.~E.} \bibnamefont{Ambrosch}},
  \bibinfo{author}{\bibfnamefont{E.~J.} \bibnamefont{Fantner}},
  \bibinfo{author}{\bibfnamefont{H.}~\bibnamefont{Clemens}}, \bibnamefont{and}
  \bibinfo{author}{\bibfnamefont{G.}~\bibnamefont{Bauer}},
  \bibinfo{journal}{Phys. Rev. B} \textbf{\bibinfo{volume}{30}},
  \bibinfo{pages}{3394} (\bibinfo{year}{1984}).

\bibitem[{\citenamefont{Ishida et~al.}(1985)\citenamefont{Ishida, Aoki, and
  Fujiyasu}}]{ishida1901}
\bibinfo{author}{\bibfnamefont{A.}~\bibnamefont{Ishida}},
  \bibinfo{author}{\bibfnamefont{M.}~\bibnamefont{Aoki}}, \bibnamefont{and}
  \bibinfo{author}{\bibfnamefont{H.}~\bibnamefont{Fujiyasu}},
  \bibinfo{journal}{Journal of Applied Physics} \textbf{\bibinfo{volume}{58}},
  \bibinfo{pages}{1901} (\bibinfo{year}{1985}).

\bibitem[{\citenamefont{Barone et~al.}(2013)\citenamefont{Barone, Rauch,
  Di~Sante, Henk, Mertig, and Picozzi}}]{PhysRevB88045207}
\bibinfo{author}{\bibfnamefont{P.}~\bibnamefont{Barone}},
  \bibinfo{author}{\bibfnamefont{T.~c.~v.} \bibnamefont{Rauch}},
  \bibinfo{author}{\bibfnamefont{D.}~\bibnamefont{Di~Sante}},
  \bibinfo{author}{\bibfnamefont{J.}~\bibnamefont{Henk}},
  \bibinfo{author}{\bibfnamefont{I.}~\bibnamefont{Mertig}}, \bibnamefont{and}
  \bibinfo{author}{\bibfnamefont{S.}~\bibnamefont{Picozzi}},
  \bibinfo{journal}{Phys. Rev. B} \textbf{\bibinfo{volume}{88}},
  \bibinfo{pages}{045207} (\bibinfo{year}{2013}).

\end{thebibliography}

\end{document}